\documentclass[a4paper]{jpconf}
\usepackage{graphicx}

\usepackage{amsmath}

\begin{document}

\title{Modelling the stellar soft-photon energy density profile of globular clusters}

\author{P L Prinsloo$^{1}$, C Venter$^{1}$, I B\"usching$^{1}$ and A Kopp$^{1,2}$}

\address{$^{1}$ Centre for Space Research, North-West University, Potchefstroom Campus, Private Bag X6001, Potchefstroom
2520, South Africa}
\address{$^{2}$ Institut f\"ur Experimentelle und Angewandte Physik, Christian-Albrechts-Universit\"at zu Kiel, Leibnizstrasse 11,
24118 Kiel, Germany}

\ead{21696764@nwu.ac.za}

\begin{abstract}
Recent observations by the $\textit{Fermi}$ Large Area Telescope (LAT) and the High Energy Stereoscopic System (H.E.S.S.) have revealed globular clusters (GC) to be sources of high-energy (HE) and very-high-energy (VHE) $\gamma$-rays. It has been suggested that the presence of large numbers of millisecond pulsars (MSPs) within these clusters may be either directly responsible for these $\gamma$-ray fluxes through emission of pulsed curvature radiation, or indirectly through the injection of relativistic leptons into the cluster. These relativistic particles are plausibly re-accelerated in shocks, created by the collision of stellar winds, before interacting with the soft-photon radiation field set up by the stellar population of the host cluster. Inverse Compton (IC) scattering then produces $\gamma$-radiation in the TeV band. In order to calculate the IC spectrum, an accurate profile for the energy density of the soft-photon field is required. We construct such a profile by deriving a radially-dependent expression for the stellar energy density, and then solve it numerically. As a next step, the average energy density values for three different regions of the cluster (demarcated by its core, half-mass, and tidal radii) are determined, which we consequently import into an existing radiation code to predict the TeV $\gamma$-ray spectrum. As an application, we consider the case of Terzan 5, boasting a population of 34 radio MSPs, and compare our predicted spectrum with that measured by H.E.S.S.
\end{abstract}

\section{Introduction}
\label{intro}
Globular clusters (GC) are large spherical collections of roughly a hundred thousand to a million gravitationally bound stars \cite{Tayler94}. The cluster cores contain the more massive stars and have very high stellar densities, creating favourable conditions for binary interaction \cite{PooleyHut06}. Expressions for the mass density profile of a GC can be constructed using Michie-King multi-mass models, for example \cite{KuranovPostnov06}

\begin{equation}
	\rho(r)= \rho_{0}
	\begin{cases}
	1 & r<r_{\rm c} \\
	(r_{\rm c}/r)^{2} & r_{\rm c}<r<r_{\rm h} \\
	(r_{\rm c}r_{\rm h})^{2}/r^{4} & r_{\rm h}<r<r_{\rm t},
	\end{cases}
	\label{massdense}	
\end{equation} 
where $r$ is the radial distance from the GC centre, $r_{\rm c}$ is the core radius, $r_{\rm h}$ the half-mass radius, $r_{\rm t}$ the tidal radius, and $\rho_{0}$ a normalisation constant. It would appear as if GCs are suitable to host large populations of millisecond pulsars (MSPs), since GCs are found to harbour a great number of stellar binary members \cite{Alpar82,CamiloRasio05}. Also, GCs are ancient objects with mean ages of 12.8$\pm$1.0 Gyr \cite{Krauss2000}, so that one would expect them to contain many evolved stellar products. In fact, the presence of 144 pulsars in 28 different clusters has been established to date\footnote{See www.naic.edu/$\sim$pfreire/GCpsr.html, doa: 1 Dec. 2014.}, although it has been predicted that there may be as many as a few hundred MSPs in cluster centres \cite{Tavani93}.   

Recent observations have revealed several GCs to be sources of HE and VHE\footnote{HE: High-energy $E >$ 100 MeV, VHE: Very-high-energy $E >$ 100 GeV \cite{Bednarek10}.} $\gamma$-radiation. The Large Area Telescope (LAT) aboard the $\textit{Fermi}$ Gamma-ray Space Telescope, for instance, detected $\gamma$-ray emission from clusters such as 47 Tucanae and Terzan~5 (Ter5) \cite{Abdo09,Abdo10,Kong10,Tam11}. Ter5 was also revealed by the ground-based High Energy Stereoscopic System (H.E.S.S.) as a source of VHE $\gamma$-radiation \cite{Abramowski11}. Since GCs are known hosts of MSPs, these ancient stars are thought to be responsible for the $\gamma$-ray fluxes. Ter5, in particular, hosts the largest known number of radio MSPs of all GCs with a population$^1$ of 34, and also boasts one of the highest central densities and stellar collision rates \cite{Lanzoni10}. It is situated at a distance of 5.9 kpc from Earth \cite{Valentietal2007}, has core, half-mass, and tidal radii of 0.15$^\prime$, 0.52$^\prime$ and 4.6$^\prime$ respectively, a total luminosity of 8$\times$10$^5 L_{\odot}$ \cite{Lanzoni10}, and is expected to harbour as many as 200 MSPs \cite{FruchterGoss2000,Kong10}.

One particular manner in which the energy of pulsars dissipates is through the ejection of particles in the form of relativistic pulsar winds \cite{ReynoldsChevalier84}. These are thought to be accelerated to relativistic speeds either within the magnetosphere of the MSP \cite{Venter09}, or in relativistic shocks caused by the collision of pulsar winds \cite{BS07}. Although such particles are expected to account for a very small fraction of a pulsar's spin-down luminosity, with an energy conversion efficiency of as little as $\eta \sim$ 0.01 \cite{VenterDeJager08}, they are plausibly responsible for emission in the TeV band \cite{Bednarek10}. The relevant radiation mechanism is thought to be inverse Compton (IC) scattering\footnote{For an alternative interpretation, see \cite{thisProceed}.} where relativistic leptons upscatter soft (or low-energy) photons to $\gamma$-rays. 
The IC-emissivity $Q_{{\rm comp},j}$, related to the scattered photon spectrum per incident electron, is \cite{Zhang08},
\begin{equation}
Q_{{\rm comp},j}(E_{\rm \gamma})=4\pi\int_{0}^{\infty}n_{j}(\epsilon,r)d\epsilon\times\int_{E_{\rm e,thresh}}^{ E_{\rm e,max}}J_{\rm e}(E_{\rm e})F(\epsilon,E_{\rm \gamma},E_{\rm e})\, dE_{\rm e},
\label{emissivity}
\end{equation} 
with $E_{\gamma}$ the upscattered photon energy, and $E_{\rm e}$ the electron energy (the limits of the rightmost integral signify the threshold and maximum electron energy).
Here the $\it J_{\rm e} $ component is a steady-state particle spectrum related to the lepton injection spectrum, which is in turn related to the pulsar spin-down luminosity, and also entails the particle transport. The $\textit{F}$ component in  equation (\ref{emissivity}) relates to the cross-section of the interaction. See~\cite{BlumenthalGould70,Zhang08} for more details. What is of interest for the purposes of this paper is the $\it n_{j}$ component, which is \text{the} photon number density, with subscript $\textit{j}$ corresponding to one of three soft-photon components (cosmic microwave background (CMB), infrared (IR) or starlight).  For a blackbody, it is given by \cite{Zhang08},
\begin{equation}
n_{j}(\epsilon)=\frac{15U_{j}}{\left(\pi kT_{j}\right)^{4}}\frac{\epsilon^{2}}{e^{\left(\epsilon/kT_{j}\right)}-1}.
\label{photondense}
\end{equation}       
Here $\epsilon$ represents the initial energy of the photon before scattering, $\textit{k}$ is the Boltzmann constant, and $\it T_{j}$ and $\it U_{j}$ respectively represent the temperature and energy density corresponding to each soft photon component. In an environment such as that created by GCs, one would expect the vast stellar populations to induce a prominent starlight component with high energy density $\textit{U}$, which decreases with increasing distance from the cluster centre, similar to the stellar population density.
 
In some models, the energy density has been represented by average values within two zones, demarcated by the cluster centre, the core and half-mass radii \cite{Venter09,VenterDeJager08}. It is the objective of this paper to construct a radially-dependent expression for the stellar soft-photon energy density of a GC (Section \ref{method}), and to solve it numerically for the case of Ter5 (Section \ref{result}). We then use this profile to predict the $\gamma$-radiation spectrum of MSPs within Ter5 (Section \ref{result}) whilst regarding IC scattering as the predominant radiation mechanism. Following this, comparisons can be made with the predicted spectra of other models (Section \ref{result}-\ref{conclude}).  

\section{Constructing the energy density profile}
\label{method} 
We assume that the stars in a GC radiate as blackbodies. By integrating the specific intensity, $I_{\nu}=(2h\nu^{3}/c^{2})(\exp(h\nu/kT)-1)^{-1}$ \cite{Tayler94} with units $\rm erg\,s^{-1}\,cm^{-2}\,Hz^{-1}\,sr^{-1}$ and $T$ the temperature, over all frequencies, one obtains the total intensity, $I = 4\sigma T^{4} /\pi$, with $\sigma=(2k^{4}\pi^{5})/(15h^{3}c^{2})$ the Stefan-Boltzmann constant, $h$ the Planck constant, and $c$ the speed of light.
The energy density due to a single star is therefore $u_{\rm s}=(4\pi/c)I$. We next scale $u_{\rm s}$ by a factor $(\bar{R}^{2}/d^{2})$ to compensate for the distance to the star ($d$) as well as its total radiating surface area ($\propto \bar{R}^{2}$, with $\bar{R}$ the average stellar radius).
By replacing $d$ with the separation distance $|\textbf{r}-\textbf{r}^{\prime}|$ between the contributing source at radius $\it r^{\prime}$ and an observer within the cluster at radius $r$ (with the GC centre at the origin), and assuming identical properties for all stars (with average stellar mass $\bar{m}\approx M_{\odot}$, and $M_{\odot}$ the solar mass), we can calculate the total energy density profile as 
\begin{equation}
	u(r)=\int u_{\rm s}N(r^{\prime})\,dr^{\prime}=\int u_{\rm s}\frac{\rho(r^{\prime})}{\bar{m}}\,dV^{\prime},
\end{equation}
where $dN = N(r^{\prime})\,dr^{\prime}$ is the number of stars in the interval $(r^{\prime},r^{\prime}+dr^{\prime})$, which is equivalent to $\rho(r^{\prime})dV^{\prime}/\bar{m}$ when assuming spherical symmetry.
Using axial symmetry, the cosine rule, and making the substitution $p(\theta)=r^{\prime 2}+r^{2}-2r^{\prime}r\cos\theta$, we finally obtain
\begin{equation}
u(r) = \frac{8\pi^{2}R^{2}}{c}\frac{I}{\bar{m}}\frac{1}{r}\int_{0}^{r_{t}}\rho(r^\prime)r^\prime\ln\left(\frac{|r^\prime+r|}{|r^\prime-r|}\right)dr^\prime.
\label{finalU}
\end{equation}      
For $\rho(r^\prime)$, we use equation (\ref{massdense}), and normalise this by setting the total cluster mass $M_{\rm total} = N_{\rm total}\bar{m} = \int\rho(r)\,dV = 4\pi\int\rho(r)r^{2}\,dr$,       
where $N_{\rm tot}=L_{\rm tot}/L_{\odot}$ is the total number of stars in the cluster. Here $L_{\rm tot}$ is the total GC luminosity, and all stars are assumed to have $L\approx L_{\odot}$ (as a first approximation;\cite{BS07}), with $L_{\odot}$ the solar luminosity. We can then express the central mass-density $\rho_{0}$ in terms of measurable cluster parameters and solve the integral in equation (\ref{finalU}) numerically.

\section{Results and discussion}
\label{result}

\begin{figure}[htbp]
	\centering
	\includegraphics[scale=0.9]{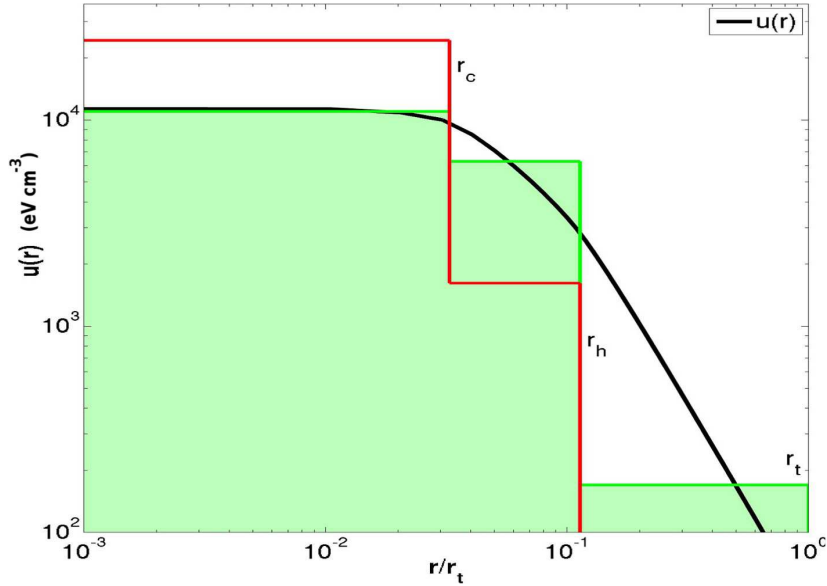}
	\caption[Comparison of energy densities for Ter5] 
	{Comparison of energy densities calculated for Ter5. The average energy densities for three regions (shown in green), demarcated by the core, half-mass and tidal radii, are calculated from the energy-density profile (solid black line) of Ter5. Shown in red are the average energy densities for two zones used by \cite{VenterDeJager08,Venter09}.    
	}
	\label{ucompare}
\end{figure} 

\begin{figure}[tbp]
	\centering
	\includegraphics[scale=0.9]{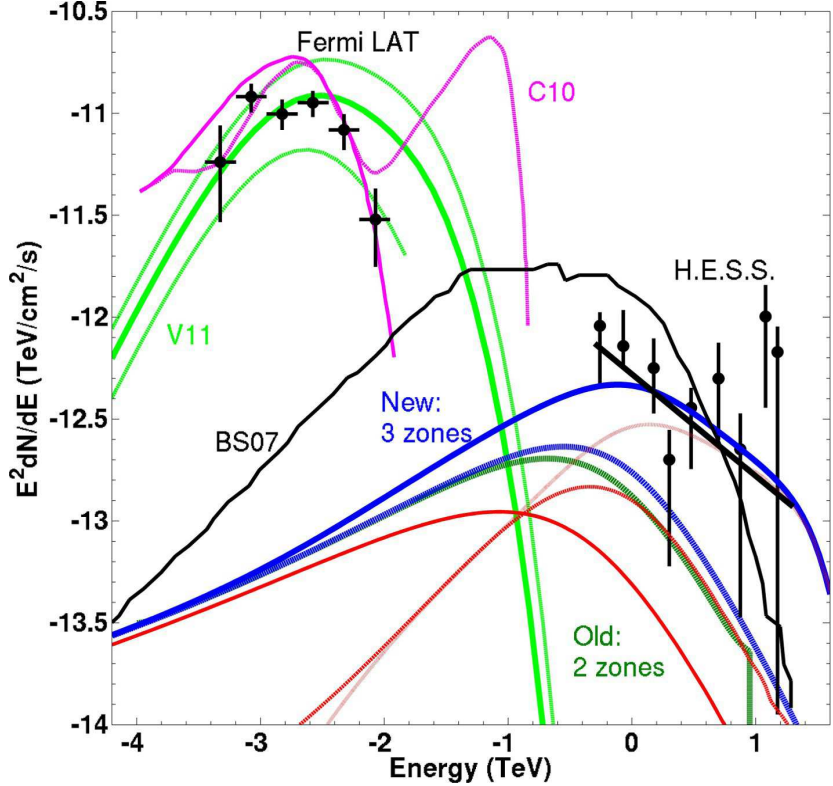}
	\caption[The Inverse Compton spectra for Terzan 5] 
	{Predicted $\gamma$-ray spectra for Ter5. Here, the contributions to the IC spectrum of the energy densities for the three zones constructed in this paper are shown in red, and their collective contribution with a solid, bright blue line. The faint blue line shows the collective contribution of the first two zones constructed in this paper. Take note that these spectra have been scaled up by a factor of three so that the resultant spectrum is better aligned with the H.E.S.S. data. Furthermore, the IC spectrum obtained by \cite{VenterDeJager08} is shown in dark green and is labelled `Old: 2 Zones', while a scaled prediction of \cite{BS07} is shown in black with label `BS07'. The lime green and magenta lines (labelled V11 and C10) are the spectra calculated for the HE band \cite{Venter11,Cheng10}.        
	}
	\label{IC_spectra}
\end{figure}

A general proportionality relation of $u \propto N_{\rm tot}R^{2}T^{4}$ can be deduced for the energy density from equation (\ref{finalU}). This effectively means that doubling either the total number of stars, the stellar radius, or the stellar temperature will result in an energy density increase of factor 2, 4 or 16 respectively. Notice that the energy-density, according to this model, shows no dependence on the mean stellar mass. This is because the central mass-density $\rho_{0}$, which was determined by normalising equation (\ref{massdense}), contains a factor $\bar{m}$ which cancels with that contained in the denominator of equation (\ref{finalU}). Furthermore, solving equation (\ref{finalU}) numerically, using the measured parameters for Ter5 and assuming solar properties for its stars, an energy density profile is obtained as shown by the solid black curve in Figure \ref{ucompare}. This energy density profile is then divided into three discrete zones \footnote{This is adequate for the current application; however, we have used the full expression when preforming detailed transport calculations involving many zones \cite{Koppetal2013}.}, as shown in Figure \ref{ucompare}. The average energy densities for each of these three zones are obtained as $1.1\times10^{4}$, $6.3\times10^{3}$, and 170 eV$\,\rm cm^{-3}$, which are consequently imported into an existing radiation code \cite{Venter09} to calculate the resulting IC spectrum. The temperature and energy density of the CMB are chosen as 2.76 K and 0.27~eV$\,\rm cm^{-3}$, and the cluster magnetic field strength as 1 $\rm \mu G$. A number of $8\times10^{5}$ stars is deduced from the total cluster luminosity, while the distance and cluster radii are taken as mentioned in Section \ref{intro}. Regarding the injection spectrum, a power law is assumed and the minimum and maximum particle energies taken as 0.1 GeV and 100 TeV respectively, with a spectral index of 1.6. Furthermore, the number of MSPs is taken as 100, the average spin-down luminosity $\langle\dot{E}\rangle$ as $2\times10^{34}\,$erg$\,\rm s^{-1}$ and the particle conversion efficiency $\eta$ as 1$\%$. Having used all of these parameters, we obtain Figure \ref{IC_spectra}.

The resultant spectrum obtained from the contribution from only two zones is indicated with a faint blue line (see Figure \ref{IC_spectra}), and does not differ much from the result by \cite{VenterDeJager08} (shown in dark green). The small difference stems from a difference in the average energy density values used. However, with the inclusion of the third zone (of which the contribution is shown with the top faint red line), the resultant spectrum (thick, solid blue line) is shaped in such a way that it can be well aligned to fit the H.E.S.S. data. For the best alignment however, it is necessary to scale the spectrum up by a factor 3, which means that if the IC-spectra depends roughly on the product $N_{\rm tot}\, N_{\rm MSP}\,\eta_{\rm p}\,\langle\dot{E}_{\rm rot}\rangle$, then the total number of stars and MSPs, the energy conversion efficiency and the spin-down luminosity must each be scaled in such a manner that a resultant increase of factor 3 occurs.
    
\section{Conclusions}
\label{conclude}
It has been argued that GCs, having a high number of stars in late evolutionary stages and a high binary encounter rate, are suitable to host large populations of MSPs. In addition, GCs have been revealed by the $\textit{Fermi}$  LAT and H.E.S.S.\ as sources of HE and VHE $\gamma$-radiation. Such $\gamma$-ray emissions have been modelled to arise from the IC scattering of stellar soft photons due to interactions with relativistic particles injected by MSPs. As part of the calculation of the IC spectrum, it was necessary to construct a radially-dependent expression for the soft photon energy density of GCs. We derived this expression analytically, normalised it, and then solved it numerically for the parameters of the GC Ter5. Having divided the profile for Ter5 into three zones (while awaiting more refined particle transport calculations which would allow us to use a greater number of zones), we calculated an average energy density for each zone and consequently used these values to predict the IC spectrum of Ter5. The resultant spectrum was compared to that generated by earlier models as well as to H.E.S.S.\ data. We found that the predicted spectrum provides a good fit to the H.E.S.S. data if scaled up by a factor 3. 
This factor may be obtained by, for example, scaling $N_{\rm tot}, N_{\rm MSP}, \eta_{\rm p}$ and $\langle\dot{E}_{\rm rot}\rangle$ each by a factor of roughly 1.3. It can be concluded that the addition of the third zone greatly improved the model's performance when comparing our IC spectrum with those of models that included fewer zones. 

Our results can be expanded in a number of ways. When calculating $u(r)$, Hertzsprung-Russel diagrams may be consulted to find a more realistic mass distribution of GC stars. Also, we may generalise $u(r)$ by considering the radii and temperatures of individual stars, and not using average values for the whole population. The energy density can be derived again for asymmetrically distributed sources, and the soft photon contribution of the Galactic background should be taken into account as well \cite{Cheng10}. In addition, the surface brightness profile implied by equation (\ref{massdense}) should be compared to the observed optical surface brightness to assess the validity of this equation. Furthermore, possible ways of improving the IC~calculation include constructing a cluster magnetic field profile and using refined transport equations for a better modelling of the evolution of the particle injection spectrum. The latter will allow the average energy densities of a greater number of zones to be imported into the radiation code without loss of stability. This might further strengthen the correspondence of the predicted IC spectrum with observational data.  

\section*{Acknowledgements}
This research is based upon work supported by the South African
National Research Foundation.

\section*{References}

\bibliography{Bibliography}

\providecommand{\newblock}{}
\begin{thebibliography}{10}
\expandafter\ifx\csname url\endcsname\relax
  \def\url#1{{\tt #1}}\fi
\expandafter\ifx\csname urlprefix\endcsname\relax\def\urlprefix{URL }\fi
\providecommand{\eprint}[2][]{\url{#2}}

\bibitem{Tayler94}
Tayler R~J 1994 {\em {The Stars: Their Structure and Evolution}\/} (Cambridge
  University Press)

\bibitem{PooleyHut06}
{Pooley} D and {Hut} P 2006 {\em {IAU Joint Discussion}\/} vol~14 p~34

\bibitem{KuranovPostnov06}
{Kuranov} A~G and {Postnov} K~A 2006 {\em Astron. Lett.\/} {\bf 32} 393--405

\bibitem{Alpar82}
{Alpar} M~A, {Cheng} A~F, {Ruderman} M~A and {Shaham} J 1982 {\em Nature\/}
  {\bf 300} 728--30

\bibitem{CamiloRasio05}
{Camilo} F and {Rasio} F~A 2005 {\em {Binary Radio Pulsars}\/} ({\em {ASP Conf.
  Ser.}\/} vol 328) ed {Rasio} F~A and {Stairs} I~H p 147

\bibitem{Krauss2000}
{Krauss} L~M 2000 {\em Physics Reports\/} {\bf 333} 33--45

\bibitem{Tavani93}
{Tavani} M 1993 {\em Astrophys. J\/} {\bf 407} 135--41

\bibitem{Bednarek10}
{Bednarek} W 2011 {\em {High-Energy Emission from Pulsars and their Systems}\/}
  ed {Torres} D~F and {Rea} N pp 185--205

\bibitem{Abdo09}
{Abdo} A~A {\em et~al.\/} 2009 {\em Science\/} {\bf 326} 1512--6

\bibitem{Abdo10}
{Abdo} A~A {\em et~al.\/} 2010 {\em Astron. Astrophys.\/} {\bf 524} A75

\bibitem{Kong10}
{Kong} A~K~H, {Hui} C~Y and {Cheng} K~S 2010 {\em Astrophys. J\/} {\bf 712}
  L36--9

\bibitem{Tam11}
{Tam} P~H~T, {Kong} A~K~H, {Hui} C~Y, {Cheng} K~S, {Li} C and {Lu} T~N 2011
  {\em Astrophys. J\/} {\bf 729} 90--7

\bibitem{Abramowski11}
{Abramowski} A {\em et~al.\/} 2011 {\em Astron. Astrophys.\/} {\bf 531} L18--22

\bibitem{Lanzoni10}
{Lanzoni} B {\em et~al.\/} 2010 {\em Astrophys. J\/} {\bf 717} 653--7

\bibitem{Valentietal2007}
{Valenti} E, {Ferraro} F~R and {Origlia} L 2007 {\em Astron. J\/} {\bf 133}
  1287--301

\bibitem{FruchterGoss2000}
{Fruchter} A~S and {Goss} W~M 2000 {\em Astrophys. J\/} {\bf 536} 865--74

\bibitem{ReynoldsChevalier84}
{Reynolds} S~P and {Chevalier} R~A 1984 {\em Astrophys. J\/} {\bf 278} 630--48

\bibitem{Venter09}
{Venter} C, {De Jager} O~C and {Clapson} A~C 2009 {\em Astrophys. J\/} {\bf
  696} L52--5

\bibitem{BS07}
{Bednarek} W and {Sitarek} J 2007 {\em Mon. Not. R. Astron. Soc.\/} {\bf 377}
  920--30

\bibitem{VenterDeJager08}
{Venter} C and {de Jager} O~C 2008 {\em {AIP Conf. Ser.}\/} vol 1085 ed
  {Aharonian} F~A {\em et~al.\/} pp 277--80

\bibitem{thisProceed}
Venter C, B{\"u}sching I, Kopp A, Clapson A~C and de~Jager O~C {$\textit{These
  proceedings}$}

\bibitem{Zhang08}
{Zhang} L, {Chen} S~B and {Fang} J 2008 {\em Astrophys. J\/} {\bf 676} 1210--7

\bibitem{BlumenthalGould70}
{Blumenthal} G~R and {Gould} R~J 1970 {\em Rev. Mod. Phys.\/} {\bf 42} 237--71

\bibitem{Koppetal2013}
{Kopp} A, {Venter} C, {B{\"u}sching} I and {de Jager} O~C 2013 {\em Astrophys.
  J\/} {\bf 779} 126--37

\bibitem{Venter11}
{Venter} C, {de Jager} O~C, {Kopp} A and {B{\"u}sching} I 2011 {\em Fermi Symp.
  Proc., eConf C110509\/} (\textit{Preprint} \eprint{arXiv:1111.1289})

\bibitem{Cheng10}
{Cheng} K~S, {Chernyshov} D~O, {Dogiel} V~A, {Hui} C~Y and {Kong} A~K~H 2010
  {\em Astrophys. J\/} {\bf 723} 1219--30

\end{thebibliography}

\end{document}